# Cost-effectiveness analysis for therapy sequence in advanced cancer: A microsimulation approach with application to metastatic prostate cancer.

Elizabeth A. Handorf, J. Robert Beck, Andres Correa, Chethan Ramamurthy, Daniel M. Geynisman

**Purpose.** Patients with advanced cancer may undergo multiple lines of treatment, switching therapies as their disease progresses. The cost-effectiveness of therapy sequence is an important policy question, but challenging to model. Motivated by a study of metastatic prostate cancer, we develop a microsimulation framework to study therapy sequence.

**Methods.** We propose a discrete-time state transition model to study two lines of anti-cancer therapy. Based on digitized published progression-free survival (PFS) and overall survival (OS) curves, we infer event types (progression or death), and estimate transition probabilities using cumulative incidence functions with competing risks. Our model incorporates within-patient dependence over time, such that response to first-line therapy informs subsequent event probabilities. Parameters governing the degree of within-patient dependence can be used to calibrate the model-based results to those of a target trial. We demonstrate these methods in a cost-effectiveness study of two therapy sequences for metastatic hormone sensitive prostate cancer, where Docetaxel (DCT) and Abiraterone Acetate (AA) are both appropriate for use in either first or second line treatment. We assess costs, Quality-Adjusted Life Years (QALYs) and Incremental Cost Effectiveness Ratio (ICER) for two treatment strategies: DCT → AA vs AA → DCT.

**Results.** Using digitized survival curves from relevant clinical trials, we identified 8.6-13.9% of PFS times that should be categorized as deaths, allowing for estimation of cumulative incidence functions. Models assuming within-patient independence overestimated OS time, corrected with our calibration approach. Correction resulted in meaningful changes in the difference in QALYs between treatment strategies (0.07 vs 0.15) and the ICER (-$76,836/QALY vs -$21,030/QALY).

**Conclusions.** Microsimulation models can be successfully used to study cost-effectiveness of therapy sequences, taking care to account correctly for within-patient dependence.

## 1. Background

In modern oncology practice, the availability of new anti-cancer therapies that improve survival leads to more patients living longer with advanced cancers. Often, such patients undergo multiple lines of therapies, switching treatments when their disease progresses. For many patients and scenarios, the clinically optimal order of therapies is not clear and rarely studied by prospective randomized trials, with new treatment paradigms continuing to evolve. In many cancers, specific agents are medically appropriate in several lines of treatment, which motivates the question: Which therapies should be given, and in what order? As new therapies are often expensive, and may provide only marginal improvements in survival and quality of life, the cost-effectiveness of different therapy sequences is highly relevant for patient and provider decision making.

Although generally under-studied, the cost-effectiveness of various therapy sequences has been examined in several advanced cancers, including BRAF wild-type melanoma, [1] EGFR mutated non-small-cell lung cancer, [2] HER2+ breast cancer, [3, 4] and KRAS wild-type colorectal cancer. [5] These cost-effectiveness analyses often use state-based Markov models. [1, 3-5] A major limitation of this approach is the "memoryless" property of the Markov process; that is, the probability of moving from one health state to another depends only on the current state. Semi-Markov models relax this assumption, but when prior health states inform future events, such models require many additional states to encode prior information and can become unwieldy (a phenomenon referred to as "state explosion"). This has forced previous studies of therapy sequence to make unrealistic simplifying assumptions.

In the therapy switching problem, several analytic challenges make modelling complex. First is the staggered start times of later lines of therapy, which can lead to outcomes and transition probabilities dependent on both the time spent in the state and the total time from the start of the model. Second, the probability of death is highly dependent on prior progression events. That is, patients are much more likely to die once their disease has progressed. Third, patients' outcomes are dependent over time. Patients who do not respond well to first-line therapy and experience early progression are more likely to do poorly in subsequent lines of treatment due to the inherent aggressive biology of their cancer or other factors like comorbid conditions. Fourth, not all first-line trial patients would be eligible for second-line trials, as some patients' overall health would deteriorate. The exclusion of such patients from second-line trials would lead to an overestimate of survival for the full first-line cohort (i.e. survivorship bias).

We propose using microsimulation models, [6] which are discrete-time state-transition models that aggregate simulated individual-level trajectories. [7-10] These models have been used successfully in a variety of health-economic applications, and are particularly useful when individual-level factors inform future transition probabilities. These models require simulating a large number of patients to ensure stability of resulting model estimates, which can make the models computationally intensive. [6] Nevertheless, modern computing solutions (better hardware and use of higher-level programming languages) have made these models more practical to run in recent years.[8] We show how we can use microsimulation models to address each of the problems listed above, which we describe in detail in the methods section.

This study is motivated by the analysis of two therapy options for metastatic prostate cancer, Abiraterone Acetate (AA) and Docetaxel (DCT). The cost-effectiveness of these two treatments has been of substantial interest, for both first-line (hormone sensitive) and second-line (hormone resistant) disease. [11-13] Although AA has a better side-effect profile, both DCT for first-line therapy followed by AA after progression, and AA followed by DCT, are consistent with the standard of care. Until very recently, AA was not available as a generic, with a list price upwards of $10,000/month, while DCT has been available as a generic drug for years. Although AA is now available off-patent, the question of therapy order is not unique to this situation. As new, expensive agents come on the market, patients, physicians, and payers will need to make value judgements when considering not just whether to give a specific therapy, but also when the therapy should be given.

## 2. Methods

### 2.1 Microsimulation models for multiple lines of therapy

**2.1.1 Model framework.** Because of the complex dependence over time, we propose a discrete time state-based microsimulation model. Figure 1 shows the general structure of such a model. Patients would start in "Line 1" therapy, and stay in that state until disease progression. At this point, they would switch therapies and move to the "Line 2" state. After progression on "Line 2" therapy, the patient would move into the "Extensive Disease" state. This refers to a condition wherein patients would have poor quality of life and require expensive care. This disease state may have active anti-cancer therapy or simply supportive therapy, depending on the disease being studied. Patients could also move directly from "Line 1" to "Extensive Disease"; the proportion of patients who do not go on to receive "Line 2" therapy would depend on disease characteristics. In prostate cancer where patients are generally healthy after first-line treatment, most patients would go on to "Line 2", but in a more aggressive disease like pancreatic cancer the probability of moving directly to "Extensive Disease" would be much higher. Patients can move from any state to "Death", which is an absorbing endpoint.

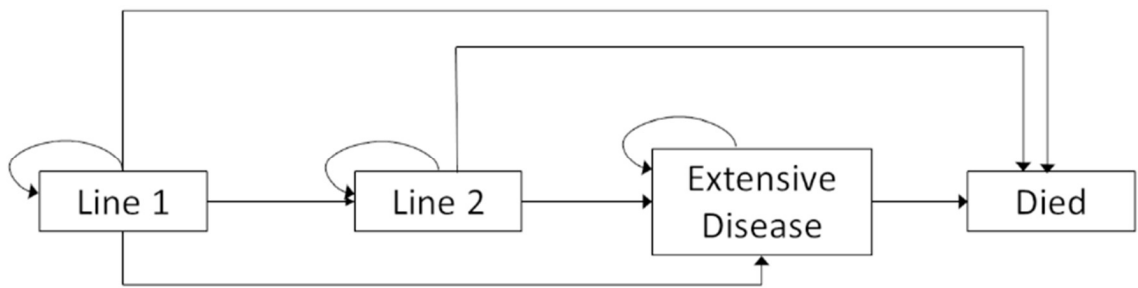

**Figure 1: General model structure for two lines of therapy**

**2.1.2 Estimation of transition probabilities.**
Many cost-effectiveness studies use published survival curves from clinical trials to inform model transition probabilities. To obtain the correct transition probabilities for this model, one must take into account the dependence of progression and death. Overall survival curves from clinical trials of first-line therapy cannot be used directly to estimate probabilities of death, as doing so would result in too many individuals dying while on therapy (without having first progressed). Although some deaths occur before progression is noted clinically, the majority will happen after progression. We propose to use a competing risks framework to estimate transition probabilities. [14] That is, we will estimate the cumulative incidence function of progression accounting for the competing risk of death.

We start the estimation process by abstracted data from published studies. Commercial software is available that allows the investigator to digitize survival curves, resulting in a dataset defining the survival step function (probabilities and times). After obtaining these results, the method proposed by Guyot et. al. allows the researcher to combine the digitized curves with published numbers at risk to infer the underlying event and censoring times. [15] After we apply this algorithm to the progression free survival (PFS) and overall survival (OS) curves, we have vectors of times ($T$) and censoring indicators ($\delta$): $T_{OS}$ and $\delta_{OS}$ for OS and $T_P$ and $\delta_P$ for PFS. For the censoring indicators, 1 indicates that an event occurred, and 0 indicates that the observation was censored.

In many clinical studies, death is considered an event when estimating progression free survival time (PFS). After digitizing the survival curves we propose an algorithm to determine which events should be classified as deaths and which should be considered progressions. One can look for close matches between death event times (abstracted from the OS curve) and progression/death event times (abstracted from the PFS curve). PFS events that occur at the same time as deaths (within some margin of error) can be assumed to be deaths. Using this logic, we can develop a vector of event type indicators ($\delta_{CR}$) that can be used in a competing risks model ($\delta_{CR}$). For $\delta_{CR}$ 0 denotes a censored observation, 1 denotes progression, and 2 denotes death.

Formally, we initially set $\delta_{CR} = \delta_P$ and $T_{CR} = T_P$.

$$\forall \{T_{OS_i} | \delta_{OS_i} = 1\},$$
$$\text{Let } A_i = \{j | T_{CR_j} \in [T_{OS_i} - \varepsilon, T_{OS_i} + \varepsilon] \,\&\, \delta_{CR_j} = 1\}$$
$$|A_i| > 0 \rightarrow \delta_{CR a_1} = 2.$$

Where $\varepsilon$ is the allowable margin of error. More intuitively, this algorithm initially defines the vector of event type indicators to be the vector of PFS censoring indicators. For each death event, it looks for all progression events within some acceptable margin. If one or more progression event meets this criteria, it sets the first of them to be a death event. Note that not all death events will have a corresponding time in the PFS curve, as most patients progress before death. After one has obtained an inferred dataset with event times and event types, cumulative incidence curves can be estimated using standard methods for competing risks.[16]

**2.1.3 Model calibration.**

For second line treatment and beyond, trial results may be too optimistic for the full cohort, as some 1st line trial-eligible patients will have deteriorated health at the time of progression (and would be excluded from most subsequent clinical trials). Also, we anticipate dependence between lines of therapy. Patients with more aggressive disease will tend to have worse response and shorter PFS in all lines of therapy.[17] This can result in a substantial overestimate of overall survival for the cohort. We propose a calibration approach that simultaneously addresses both of these issues.

We can increase the event probabilities in Line 2 and Extensive Disease states when time in 1st line ($T_1$) is short by applying a hazard ratio (HR) to the survival curves, where the HR is some non-increasing function of T1. The function defining the HR is denoted as G(T1).

This will result in a correlation between the time spent in "Line 1" and "Line 2" states. Further, if G(t)>1 for all or most values of T1, then it can correct the overestimate of survival outcomes for later lines of therapy. One simple and easy to interpret functional form is a linear decrease in G(t) from some maximum HR (θ) to a HR of 1 at some maximum time (ω).

$$G(T_{1i}) = \begin{cases} \theta - \dfrac{(\theta - 1)}{\omega} T_{1i} & T_{1i} < \omega \\ 1 & T_{1i} \geq \omega \end{cases}$$

Where $T_{1i}$ = Time spent by subject *i* in Line 1 before first disease progression, θ = Largest hazard ratio applied to any subject, and ω = Smallest time spent in Line 1 for which no penalty will be applied for later lines of therapy.

We propose using a bounded Nelder-Mead algorithm to choose parameters (e.g. θ and ω) that best calibrate the OS curve to the trial data. This is a simplex method that uses gradient descent to find parameters which minimize the objective function. [18, 19] Nelder-Mead is particularly useful here because it does not require a fully defined parametric function and can optimize multiple parameters in a constrained parameter space. We propose minimizing the sum of squared error (SSE) between the microsimulation model-based OS curve, $S_{mod}(t)$ and the target OS curve from the relevant trial, $S_{tgt}(t)$. As we use a discrete time model, a natural choice is to calculate the SSE based on the difference between the two survival curves at the start of each cycle. We use the following cost function:

$$F(\theta, \omega | T_1) = \sum_{c=1}^{C} \left(S_{mod}(t = c | \theta, \omega, T_1) - S_{tgt}(t = c)\right)^2$$

Where C is the total number of cycles. We then find

$$\arg\min_{\theta, \omega} F(\theta, \omega | T_1) \text{ subject to } \theta \in [1, \infty), \omega \in [1, \infty)$$

We can also use bounds to restrict the range of the final parameter estimates to clinically plausible values. After identifying the best values, the resulting model-based survival curves can be tested for lack of fit against the target survival curves using the Komolgorov-Smirnov test.

## 2.2 Prostate Cancer Application

### 2.2.1 Background

These modeling methods were motivated by a study of metastatic hormone-sensitive prostate cancer (mHSPC). This refers to de-novo metastatic prostate cancer (patients initially diagnosed with M1 disease), or disease refractory to local treatment not previously treated with hormonal therapy (i.e. those treated with surgery or radiation therapy). For decades, the accepted first-line treatment for mHSPC was Androgen Deprivation Therapy (ADT) alone. After progression, patients are said to have metastatic castrate resistant prostate cancer (mCRPC), and other agents would be added to ADT. Docetaxel (DCT) is a cytotoxic chemotherapy which has long been a standard therapy for mCRPC.  Aberaterone acetate (AA) is a next generation anti-androgen therapy which has been approved in mCRPC since 2011, and has a better safety profile than cytotoxic chemotherapies. [20] Starting in 2015, two seminal phase III trials changed the treatment paradigm for mHSPC.  The CHAARTED trial demonstrated a Progression-free survival (PFS) and Overall Survival (OS) benefit for DCT+ADT vs ADT alone. [21] Next, in 2017 the LATITUDE trial also demonstrated PFS and OS benefits for AA + ADT vs. ADT alone.  [20]

First-line docetaxel and abiraterone have never been compared head-to-head, and such trials are not likely to occur.  Meta-analyses demonstrate a potential PFS benefit for AA, but OS between the two treatments is largely similar. [22] According the NCCN guidelines (version 3.2022) either AA+ADT or DCT+ADT are considered category 1 treatments (i.e. those with the strongest level of evidence) for mHSPC. Additionally, in patients who progress after either treatment, switching to the other is clinically appropriate.  Therefore, treatment sequences of either AA followed by DCT (AA→DCT) or DCT followed by AA (DCT→AA) are appropriate according to consensus guidelines (for simplicity we drop "+ADT" going forward, but note that all patients will continue to receive ADT in addition to other therapies).   In this case where overall survival outcomes are expected to be similar, the cost-effectiveness of either treatment strategy is of interest for payers and policymakers.  We use this clinical example to illustrate our microsimulation methods.

### 2.2.2 Model structure and inputs

Figure 2 shows the prostate cancer model, which is adapted from the model shown in Figure 1.  We allow for adverse events to occur after either line 1 or line 2 therapy.  We also allow patients to stop active treatment without yet experiencing disease progression; this is part of the planned course of treatment for DCT (6 cycles on Line 1, 10 cycles on Line 2).  Although AA should ideally be taken until disease progression, some patients may stop early (as was seen in the trial). Patients off therapy but without disease progression are in the "Post-Line1" or "Post-Line 2" state.

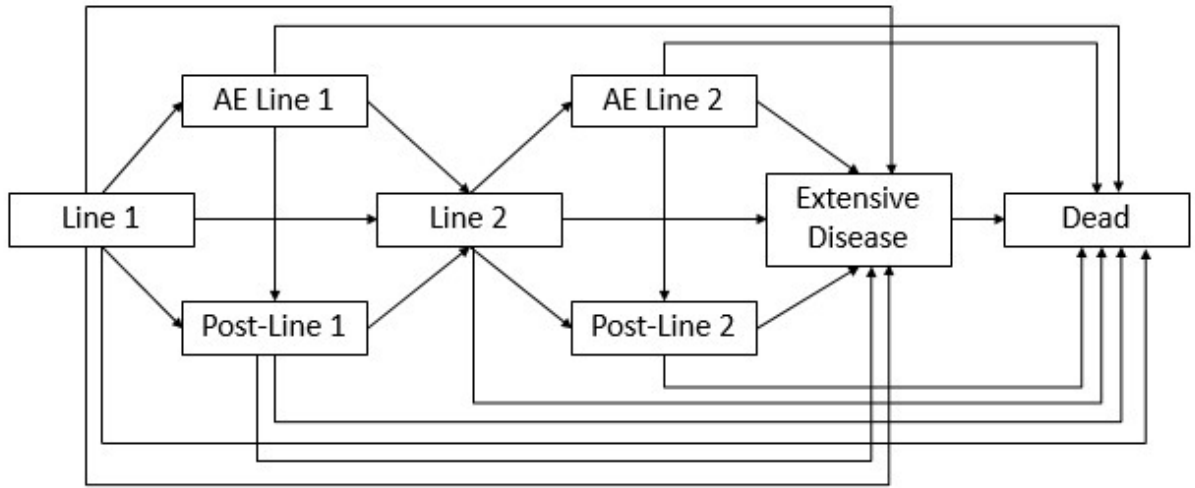

**Figure 2: States and allowable transitions in prostate cancer model**

Transition probabilities between Line 1, Line 2, and Extensive Disease were defined using the methods described above. Survival times were first extracted from digitized survival curves from relevant clinical trials (CHAARTED, LATITUDE, MANSAIL, and COU-AA-302.) [20, 21, 23, 24] Calibration parameters were then determined, and the adjusted survival curves defined the transition probabilities. Probability of death from extensive disease was based on the PROSILECA trial (which enrolled heavily pre-treated patients) [25], and these probabilities were also adjusted using the calibration methods. DigitizeIt software (DigitizeIt version 2.0, Braunschweig Germany) was used to numerically define the survival curves. We assumed a low (10%) probability of transitioning directly from Line 1 to Extensive Disease based on clinical expertise, due to the relatively healthy population at baseline and the typical course of this disease.

Other model inputs were similar to prior studies of first-line treatment of mHSPC. [11] Selected model inputs are shown in Table 1, with all model inputs and relevant assumptions listed in the supplementary material. Adverse events included in the model were those which were relatively common, and likely to substantially effect cost and/or quality of life. These included fatigue and neutropenia with and without fever. Only grade 3+ adverse events were considered, and the probabilities of these events were taken from the respective clinical trials. [20, 21, 23, 24] We assumed that afebrile neutropenia was treated with filgrastim, whereas febrile neutropenia required hospitalization.

Our model included costs of medications, medication administration, physician office visits, and hospitalizations. Medication costs were obtained from retail pharmacies,[26] or by average wholesale prices, adjusted to reflect anticipated discounting. [27] [28]. Costs of medication administration and office visits were taken from the Centers for Medicare and Medicaid Services physician fee schedule. [29] The expense of treating patients in the Extensive Disease state was estimated using results of a prior cost-effectiveness study of third-line treatments for metastatic prostate cancer (cabazitaxel after treatment with docetaxel and androgen inhibitor). [30] Costs of hospitalization for febrile neutropenia were taken from prior literature. [31] All costs were calculated in 2021 dollars. [32]

Quality of life adjustments were made via utilities. These were taken from relevant literature. [27, 33-35] Where possible, we used quality of life data from the clinical trials of interest. [36] Utilities were lower when on treatment with DCT than when on treatment with AA. We assumed that half of the patients on line 2 therapy would have symptomatic disease (e.g. bone metastasis or other symptoms which reduce quality of life). We assumed that all patients in the Extensive Disease state would have cancer symptoms, and a poorer quality of life. We used a 5-year time horizon, as this was the longest follow-up supported by trial data. Cost and effect outcomes were discounted at 3% per year. This model used a payer perspective in the United States of America. Cycle length was 3 weeks (based on DCT treatment schedule).

**Table 1: Selected model inputs other than calibrated state transitions**

| Costs (per 3 week cycle) | Value | Reference |
|---|---|---|
| AA - 2021 branded | $2396 | [26] |
| AA - 2021 generic | $296 | [26] |
| AA - on patent | $6560 | [28] |
| DCT | $2,388 | [28] |
| Treatment of Febrile Neutropenia | $19,675 | [31] |
| **Utility Values** | **Value** | **Reference** |
| L1 SD on AA | 0.83 | [36] |
| L1 SD on DCT | 0.78 | [33] |
| L2 SD on AA | 0.725 | [27, 36] |
| L2 SD on DCT | 0.675 | [27, 33] |
| Extensive disease | 0.62 | [27] |
| **Adverse Event Probabilities** | **Value** | **Reference** |
| L1 Fatigue (AA) | 0.02 | [23] |
| L1 Neutropenia (DCT) | 0.12 | [21] |
| L1 Febrile Neutropenia (DCT) | 0.061 | [21] |

### 2.2.3 Software
All analyses were performed using R software version 3.6. The code framework for the microsimulation model was adapted from Krijkamp et al. [8] Software to infer the survival times was published by Guoyt et al. [15] Cumulative incidence functions were fit using the R package cmprsk. [37] Nelder-Mead optimization was performed using the package neldermead [38].

## 3. Results

We present the results of our microsimulation approach in the context of our prostate cancer application. First, we used the method described in section 2.1.2 to separately identify progression and death events. For line 1 DCT, there were a total of 397 observations. Of these 181 were progression events. For this study, death was censored in the calculation of the PFS curves. Our method recategorized 38 of the 216 censored observations as deaths while on line 1 therapy, so overall, 9.6% of observations were identified as death events. For line 1 AA, there were a total of 597 observations, and 83 of the 247 progression events were recategorized as deaths (for this trial, death was considered a progression event). Overall 13.9% of AA progression events were recategorized. When analyzing line 2 progression-free survival curves, 45/526 (8.6%) DCT observations were categorized as deaths, and 60/564 (10.6%) AA observations were categorized as deaths.

After using our optimization approach, we identified (θ,ω) = (2.21,87) as the best calibration parameters when DCT is used as the line 1 treatment, and (5.07,36) when AA is the line 1 treatment. This indicates that the DCT survival curve best matches the trial data with a modest to small hazard ratio applied to Line 2 and Extensive Disease states, and at least some small survival penalty applied to all patients in the simulation (87 total cycles). Contrary to this, for AA, a very large hazard ratio is applied to subjects with very short times on line 1 therapy, but no additional hazards are applied to patients who are progression-free on line 1 therapy for greater than 36 cycles (9 months). The results of applying these calibration parameters are shown in Figure 3 and Table 2. With no calibration, the model substantially overestimates survival, but the calibration approach is very successful at improving fit. In spite of their large differences, interestingly, both sets of optimal parameters resulted in a non-significant test for lack of fit for both DCT and AA OS curves. Because of this, and because of the longer follow up from the DCT trial, in our clinical example, we used the DCT optimal parameters as our base case.

Table 2.

|  | DCT first | | | AA first | | |
|---|---|---|---|---|---|---|
|  | 3 year | 5 year | p-val* | 3 year | 5 year | p-val* |
| Trial data | 0.697 | 0.443 |  | 0.658 | not available |  |
| No calibration | 0.789 | 0.613 | 0.00051 | 0.776 | 0.595 | 0.0023 |
| DCT optimal parameter | 0.681 | 0.487 | 0.86 | 0.696 | 0.509 | 0.80 |
| AA optimal parameter | 0.630 | 0.495 | 0.29 | 0.664 | 0.531 | 0.99 |
| *p-val testing the model-based results to the true trial results using the Komolgorov-Smirnov test for lack of fit | | | | | | |

**Figure 3: Calibration of model-based curves to trial-based overall survival**

A. No calibration

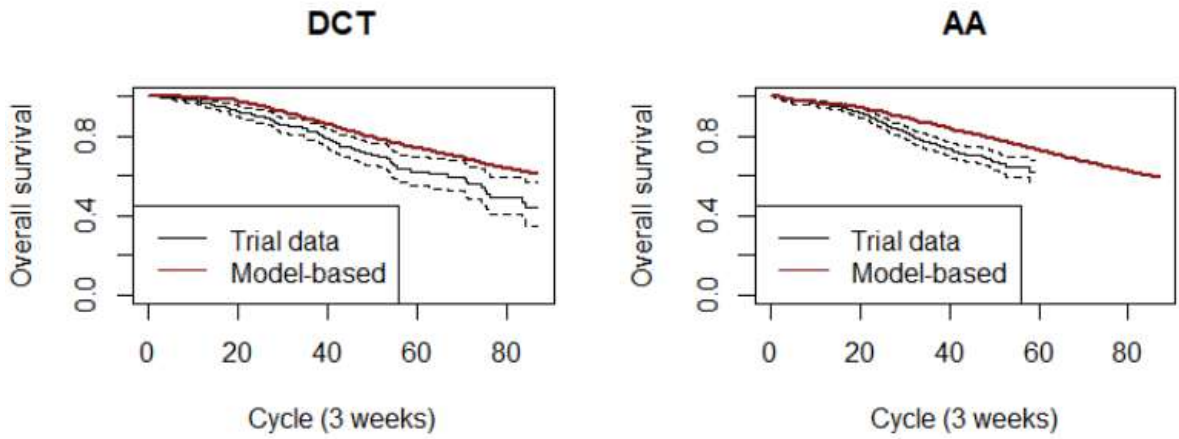

B. Optimal calibration parameters for docetaxel

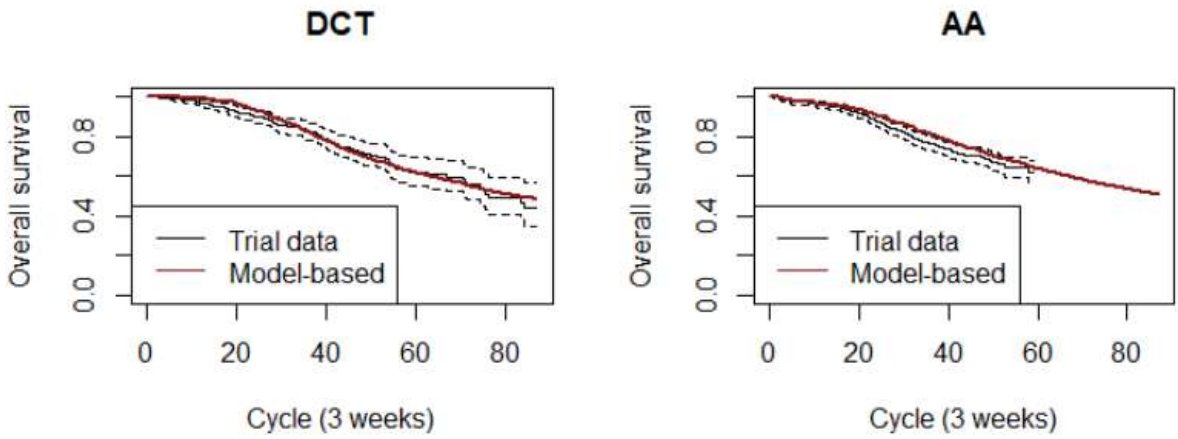

C. Optimal calibration parameters for AA

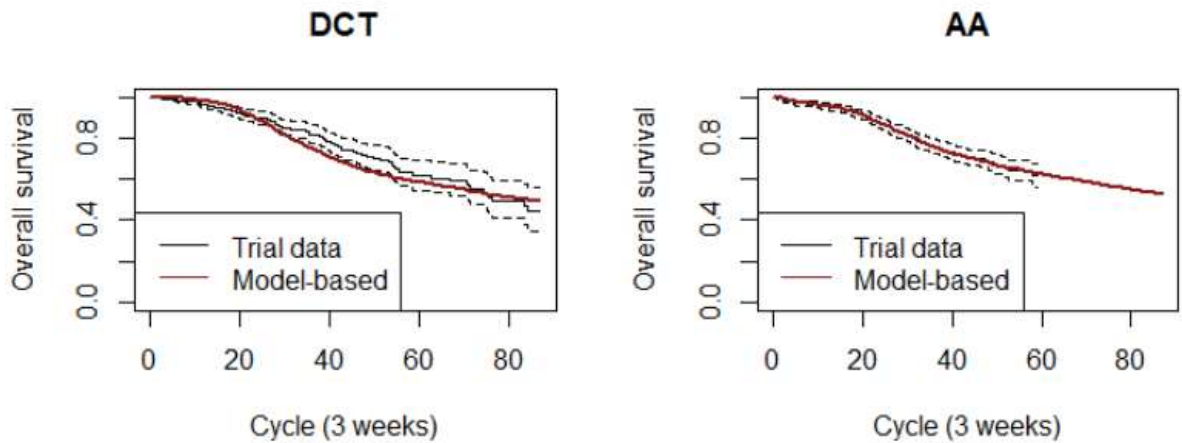

Next, we evaluated the cost-effectiveness of the two treatment strategies. We assessed how estimates changed when we varied the calibration parameters. We calculated outcomes using: 1) the optimal DCT parameters, 3) the optimal AA parameters, and 3) the DCT parameters for DCT survival probabilities, and AA parameters for AA survival probabilities (i.e. optimal parameters applied to each respective treatment). We compare these results to those of no correction. The results of the models are show in Table 3A. We see that the uncalibrated model gives the highest estimates for both costs and QALYs for both treatment strategies. The difference in QALYs was the smallest for the uncalibrated model, and was larger after corrections, particularly when applying the same correction to both strategies. Although costs were lower in the calibrated models, the difference in costs varied. When applying the optimal AA correction, the two treatment strategies had very similar costs, while the largest difference was found when applying different optimal corrections to the respective treatment strategies.

Finally, we varied the cost of AA, as it has recently become available as a generic, and the cost-effectiveness of AA in our prior analyses showed that the ICER was highly sensitive to AA cost. [11] Table 3B shows the results of these model, using the optimal calibration parameters for DCT. (As discussed above, we chose to use the optimal DCT parameters for analyses when varying AA cost). We see that, consistent with other studies, the on-patent pricing results in the AA→DCT strategy being not cost-effective, with an ICER >$1.5 million. Using current brand name pricing, AA→DCT is still not cost-effective, however, with generic pricing this strategy becomes cost-saving, and dominates the DCT→AA strategy, with an ICER of -$21,030.

Table 3: Cost-effectiveness model results
A. Effect of varying calibration parameters on model results (AA branded cost)

| Calibration Parameters | Strategy | Cost | QALY | Δ Cost | Δ QALY | ICER |
|---|---|---|---|---|---|---|
| No correction (θ,ω) = (1,1) | DCT First | $132,144 | 2.96 | | | |
| | AA First | $126,784 | 3.03 | -$5,360 | 0.07 | -$76,837 |
| DCT optimal (θ,ω) = (2.21,87) | DCT First | $119,458 | 2.72 | | | |
| | AA First | $116,392 | 2.87 | -$3,066 | 0.15 | -$21,030 |
| AA optimal (θ,ω) = (5.07,36) | DCT First | $112,393 | 2.63 | | | |
| | AA First | $111,577 | 2.82 | -$816 | 0.19 | -$4,270 |
| AA: (θ,ω) = (2.21,87) DCT: (θ,ω) = (5.07,36) | DCT First | $119,458 | 2.72 | | | |
| | AA First | $111,577 | 2.82 | -$7,881 | 0.10 | -$76,422 |

B. Effect of varying AA cost on model results (DCT optimal calibration (θ,ω) = (2.21,87))

| AA cost | Strategy | Cost | QALY | Δ Cost | Δ QALY | ICER |
|---|---|---|---|---|---|---|
| Generic (2021) | DCT First | $119,458 | 2.72 | | | |
| | AA First | $116,392 | 2.87 | -$3,066 | 0.15 | -$21,030 |
| Branded (2021) | DCT First | $136,849 | 2.72 | NA | NA | NA |
| | AA First | $211,615 | 2.87 | $74,766 | 0.15 | $512,794 |
| On-patent | DCT First | $171,330 | 2.72 | NA | NA | NA |
| | AA First | $400,412 | 2.87 | $229,082 | 0.15 | $1,571,189 |

## 4. Limitations

Our proposed methods have several limitations. One drawback of the microsimulation approach in general is its use of discrete time to approximate an underlying continuous process. This can be problematic for several reasons, as discussed by Graves et al. [39] They recommend using Discrete Event Simulation models with continuous time as an alternative. Adapting our framework to a DES model is an important area for future research. However, we note that the ability of the microsimulation model to closely match overall survival results from relevant trials demonstrates its utility in our clinical scenario.

Our calibration approach used a simple functional form for the correction. The linear correction is unlikely to be exactly correct, however, it has the benefits of being simple to estimate and interpret. Other functional forms could be considered, and any monotonic decreasing function of the hazard ratio would result in a positive correlation between time spent on line 1 and line 2 therapies. One could consider a flexible spline-based approach; however, one would need to exercise caution to avoid model overfitting. Additionally, each iteration is time consuming to fit, so models with more than 2 parameters may become computationally burdensome. In our clinical example, the linear correction was sufficient to provide good calibration.

As the focus of this manuscript is methods development, we did not perform a full sensitivity analysis to vary each parameter used in the illustrative cost-effectiveness example. Instead, we limited the sensitivity analysis to the most pertinent parameters: those used in the calibration. As the price of AA has recently changed substantially, we varied its cost as well. A full cost-effectiveness analysis of modern treatments for mHSPC is of interest for future studies, as the treatment paradigms for this disease continue to evolve. [40, 41]

## 5. Conclusions

We have developed a microsimulation strategy that successfully models treatment sequence across multiple lines of therapy. Our method allows for event probabilities to be correlated within patients across study states, and can calibrate model-based results to those of target trials. Our model could be adapted to many studies of treatment sequence. Multiple lines of therapy are very common in advanced cancers, but our work is also broadly applicable to other progressive diseases, such as chronic infections, rheumatic disease, or cardiovascular diseases.

When we first looked at our clinical example of metastatic prostate cancer, we tried to avoid incorporating dependence within patients by using the overall survival curves directly to define probabilities of death in a traditional Markov model. However, this led to 32% of patients dying while on Line 1 therapy, which is clinically unrealistic in this relatively healthy population of mHSPC patients.

Using the competing risks framework solved this problem, but led to calibration issues. One alternative approach we tried was to increase the probability that patients would move directly from Line 1 to Extensive Disease, however, in order to calibrate the model results to the OS curves, this required that an unrealistically high proportion of patients forego second line treatment. Although this is a metastatic cancer population, prostate cancer is still relatively indolent and patients may survive for years with their disease.  Further, we recognized that patients with poor outcomes on line 1 therapy are more likely to have worse responses to other therapies, due to the underlying nature of their disease.  Our calibration strategy reflects this clinical reality.  Our approach, using competing risks estimates with re-calibrated transition probabilities, enabled us to create a model with higher fidelity to real patient experiences.

In our clinical example, we show how a lack of calibration can substantially alter model estimates, in this case, by underestimating the benefit of AA→DCT over DCT→AA (0.07 QALYs without calibration vs 0.10-0.19 QALYS with calibration). Finally, we showed preliminary estimates of the effect that generic pricing will have on the cost effectiveness of these two treatment strategies.  With on patent or current branded pricing, the AA→DCT strategy is generally not cost effective.  These results are generally consistent with prior studies of line 1 therapy (where AA is found to not be cost-effective)[11, 12] although we note that here we are testing the full planned treatment strategy, not just the choice of first-line treatment. However, these results changed dramatically with the current pricing for generic DCT, and the AA→DCT strategy dominates DCT→AA, being both more effective and less costly. These results are highly sensitive to the cost of AA.  Here we used prices from one retail pharmacy; future research should obtain more comprehensive estimates. The average wholesale price of generic AA ranges from $510-$11,665 per month,[28] indicating substantial variability in costs.  A prior study in Canada estimated that generic AA would cost

$2370(CAD); such large differences in pricing will have substantial impacts on the cost-effectiveness.

Supplemental Tables

Table A1 C/E model inputs

| Variable | Value | Reference | Notes/Assumptions |
|---|---|---|---|
| Costs in $ 2021 US – per 3 week cycle | | | |
| AA - 2021 branded | 2396 | [26] | 21 days: 100mg abiraterone acetate + 5mg prednisone |
| AA - 2021 generic | 296 | [26] | 21 days: 100mg abiraterone acetate + 5mg prednisone |
| AA - on patent | 6560 | [28] | 21 days: 100mg abiraterone acetate + 5mg prednisone |
| AA non-drug costs (per cycle)$^\Delta$ | 39 | [29] | Every 3 months: office visits (CPT99214) labs: CBC, PSA, CMP (CPT codes 85025,84154,80053) |
| DCT (per 3 wk cycle) | 2386 | [28] | Dosage is 75mg/m$^2$, Assume body surface area 2.1 m$^2$ |
| DCT non-drug costs (per cycle)$^\Delta$ | 287 | [29] | Every 3 weeks: chemo admin (CPT96413), office visits (CPT99214),CBC (CPT85025) |
| | 39 | [29] | office visits every 3 months (CPT99214), labs CBC, PSA, CMP (CPT codes 85025,84154,80053) |
| ADT | 1153 | [28, 29] | Average cost per cycle drug (22.5mg) + administration (CPT96402) |
| Treatment of Neutropenia$^\$$ | 8143 | [28] | 7.35mg peg-filgrastim |
| Treatment of Febrile neutropenia | 19675 | [31, 32] | Estimate from prior study updated to 2021 dollars |
| Treatment for extensive disease | 5477 | [32] | Recent cost effectiveness study of cabazitaxel for patients previously treated with docetaxel and abiraterone or enzalutamide (estimate $94,935/LY from reported total costs, QALYs, and utilities) |
| **Utility Values** | | | |
| L1 SD on AA | 0.83 | [36] | Patient reported outcomes from LATITUDE trial |
| L1 SD on DCT | 0.78 | [33] | Patient reported outcomes on docetaxel for mHSPC |
| L1 Post-DCT recovery state | 0.78–0.83 | [33] | Assume side effects from chemotherapy resolve over 9 3-week cycles (linear increase in utility) |
| L1 Fatigue | 0.78 | [34] | Utility decrement of 0.05 for living with minor side effect |
| L1 Febrile Neutropenia | 0.47 | [35] | No utility decrement for asymptomatic neutropenia, only febrile neutropenia. |
| L2 SD on AA | 0.725 | [27, 36] | Assume half of patients have developed substantial symptoms (i.e. 50% have symptoms similar to extensive disease state) |
| L2 SD on DCT | 0.675 | [27, 33] | Assume uniform decrement from L1 to L2 (AA L1 - AA L2=0.105) |
| L2 Post-DCT recovery state | 0.675 -0.725 | [27, 33] | Assume uniform decrement from L1 to L2 (AA L1 - AA L2=0.105) |
| L2 Fatigue | 0.675 | [27, 34, 36] | Assume uniform decrement from L1 to L2 (AA L1 - AA L2=0.105) |
| L2 Febrile Neutropenia | 0.365 | [27, 35] | Assume uniform decrement from L1 to L2 (AA L1 - AA L2=0.105) |
| Progressed Disease | 0.62 | [27] | |
| Death | 0 | | |
| **AE rates** | | | |
| L1 DCT early discontinuation | 0.043-0.019 | [21] | CHAARTED trial. Per cycle rate of early discontinuation. |
| L1 AA discontinuation | 0.12 | [23] | LATITUDE trial. Overall rate of discontinuation, assume would occur in first 9 cycles. |
| L1 Fatigue (AA) | 0.02 | [23] | LATITUDE trial. Overall rate of fatigue, assume it would occur in first 9 cycles. |
| L1 Neutropenia (DCT) | 0.12 | [21] | CHAARTED trial. Overall rate of nonfebrile neutropenia (any time in first 6 cycles) |
| L1 Febrile Neutropenia (DCT) | 0.061 | [21] | CHAARTED trial. Overall rate of febrile neutropenia (any time in first 6 cycles) |
| L2 DCT early discontinuation | 0.016 | [24] | MAINSAIL trial. Per cycle rate of early discontinuation. |
| L2 AA discontinuation | 0.1 | [20] | COU-AA-302 trial. Overall rate of discontinuation, assume would occur in first 9 cycles. |
| L2 Fatigue (AA) | 0.02 | [20] | COU-AA-302 trial. Overall rate of discontinuation, assume would occur in first 9 cycles. |
| L2 Neutropenia (DCT) | 0.163 | [24] | MAINSAIL trial. Overall rate of nonfebrile neutropenia (Although they get 10 cycles, assume it would happen during the first 6 cycles) |
| L2 Febrile Neutropenia (DCT) | 0.044 | [24] | MAINSAIL trial. Overall rate of nfebrile neutropenia (Although they get 10 cycles, assume it would happen during the first 6 cycles) |